# Designing a Serious Game: Teaching Developers to Embed Privacy into Software Systems

Nalin Asanka Gamagedara Arachchilage
La Trobe University, Australia
nalin.asanka@latrobe.edu.au

Mumtaz Abdul Hameed
Technovation Consulting & Training PVT
mumtazabdulhameed@gmail.com

## Abstract

Software applications continue to challenge user privacy when users interact with them. Privacy practices (e.g. Data Minimisation (DM), Privacy by Design (PbD) or General Data Protection Regulation (GDPR)) and related "privacy engineering" methodologies exist and provide clear instructions for developers to implement privacy into software systems they develop that preserve user privacy. However, those practices and methodologies are not yet a common practice in the software development community. There has been no previous research focused on developing "educational" interventions such as serious games to enhance software developers' coding behaviour. Therefore, this research proposes a game design framework as an educational tool for software developers to improve (secure) coding behaviour, so they can develop privacy-preserving software applications that people can use. The elements of the proposed framework were incorporated into a gaming application scenario that enhances the software developers' coding behaviour through their motivation. The proposed work not only enables the development of privacy-preserving software systems but also helping the software development community to put privacy guidelines and engineering methodologies into practice.

## 1 Introduction

Companies such as Google or Facebook create a strange digital copy of individuals' lives, that they do not even know about [6]. In return, users may receive some services for free. For example, the driver tracker app called "AAMI App" [1] provides drivers with roadside assistance. However, with the app, detailed GPS logs not only reveal where she travels but how fast she drives, which route she takes, which ATMs she stops at, and also what medical clinics she has visited. At a glance, these are security flaws in software applications, so do the privacy breaches on a different lens when looking at from the user's point of view. Therefore, software applications should be developed preserving user privacy in mind rather than developing for the sake of it.

Hackers are interested in humans' psychological flaws, and therefore, targeting humans is the first thing when leveraging their attack [3]. One can argue that it is almost impossible to eliminate the end-user from using software systems, thus it requires manual, human input or interaction sometimes. On the other hand, one cannot deny the fact that incidents like Cambridge Analytica, perpetrators are targeting end-users and manipulating them through modelling their behaviour to leverage privacy breaches [6]. Therefore, it could be argued that software creators (e.g. designers or developers) are also responsible to preserve the privacy of users through the systems they develop. Nevertheless, users are unaware and find it quite difficult to understand how their data is collected, processed, stored and shared in online social networks, such as Facebook or Twitter. Therefore, software systems that access, process, store and share user data should be implemented with "privacy in mind", so that end-users' privacy may not compromise when interacting with these systems [7].

For this, software creators are expected to "humanly" design and develop software systems that people can better interact with and preserve their privacy [17, 18]. There are a number of privacy practices that are well established and widely known in the literature to guide software developers to embed privacy into software systems. For instance, Privacy by Design (PbD) [7], Fair Information Practices (FIP) [24] and Data Minimization (DM) [18]. PbD is a process of building privacy into the software design specifications, architectures and processes [7]. To build privacy in, it is imperative to understand the privacy impacts (a.k.a. Privacy Impact Assessment (PIA) [16]). When developing software applications, DM principle describes that data should only be collected if they are related to the purpose of the application, and should be processed only for the purpose which they were collected [18]. Moreover, FIP states that users should have access and control over their data even after they disclose data into a system [24]. Soon after the Cambridge Analytica incident, a new EU legislation (i.e. General Data Protection Regulation (GDPR)) came into force introducing a set guideline for the collection and processing of personal information from individuals [25].

[1]https://www.aami.com.au/app.html

However, it is still an open question whether or not software creators follow these practices/principles when designing and developing software systems that preserve user privacy. On the other hand, software developers may need help too (e.g. tooling support or training), as they are neither privacy nor security experts in most cases [27]. Traditionally, they are task-oriented [28], which may hinder when asking them to embed privacy into software systems they develop. Therefore, facilitating software developers with methodologies (e.g. a step by step approach to implement DM, FIP or DM) or providing them with educational/training interventions to embed privacy into software systems they develop, will enable them to develop privacy-preserving software systems [22]. Otherwise, it could be argued that how end-users protect their privacy, if software applications they interact with are not being designed and developed taking privacy into consideration. Previous research has developed a systematic approach (i.e. methodology), enabling developers to embed privacy into software systems through data using Data Minimisation principle (DM) [21]. DM privacy concept focuses on how data relates to privacy and controlling the use of data in systems helps implementing privacy [11]. Authors have revealed through an empirical investigation their developed methodology encourages developers to embed privacy into software systems. However, the challenge is to put this existing methodology into practice [21], so developers can embed privacy into software systems they develop. **Therefore, this research contributes to develop- ing a serious game design framework (Figure 1) as an educational intervention to enhance developers' cod- ing behaviour through motivation**. So, they can develop privacy-preserving software systems that people can use. To achieve this, we derive elements from a previously invented game design framework [2], Bloom's taxonomy [12] and data minimisation model [21].

The remainder of this paper is structured as follows. Section 2 describes the background of the study. Section 3 proposes a game design framework as an educational tool to enhance software developers' coding behaviour. Then the proposed game design framework is applied to a possible application scenario in Section 4. Section 5 provides conclusions and opens up opportunities for future work that may extend the current research presented in this paper.

## 2 Background

Software creators' engagement towards resulting privacy in software systems has become an interest of privacy researchers recently [26]. So far, however, several recent studies have investigated developers' issues, perception, engagement and interaction with privacy requirements when developing software systems that preserve user privacy [5, 8, 19, 26]. In addition, previous research has also discussed privacy practices from the organisational point of view [9].

Hadar et al. [9] have reported that organisational culture and policies play a significant role (i.e. both positively and negatively) in helping developers to consider privacy when developing software applications. Likewise, both Sheth et al. [23] and Jain et al. [10] stressed the importance of having a set of policies and guidelines in place within organisations that can guide software developers to successfully implement privacy into software applications. Nevertheless, so far too little attention has been paid to investigate issues that need to be addressed in privacy policies and how to set them up within the organisational context. This enables organisations to effectively guide developers to embed privacy into software systems that preserve user privacy.

Focusing on the engagement of software developers in developing privacy, Ayalon et al. [4] explain that developers do not follow privacy guidelines unless there is an existing methodological framework in use. Similarly, Sheth et al. [23] have revealed through an empirical investigation that developers find it difficult to understand privacy requirements and implement them into software systems. In the same vain, Oetzel et al. [16] demonstrate that developers require significant effort to estimate privacy risks from a user perspective. Furthermore, Ayalon et al. and Sheth et al. [4, 23] discuss that developers encounter problems and have difficulties when attempting to embed privacy into software systems they develop. Overall, these studies highlight the need for a systematic approach for guiding software developers to embed privacy into software applications that preserve user privacy [10].

Previous research has developed a Privacy Engineering Methodology (PEM), a step-by-step approach, guiding developers to implement privacy into software systems through understanding data (i.e. using DM) [21]. Their study investigated the impact of five constructs (derived from the Technology Acceptance Model(TAM)) on the software developers' intention to follow the developed methodology (i.e. PEM). The study findings revealed that the developed methodology encourages software developers to embed privacy into software systems that preserve user privacy [22]. Furthermore, authors have suggested that getting their developed methodology implemented into an interactive platform, for example, a gaming platform, where developers can better interact with and learn to enhance their coding behaviour. So, they can develop privacy preserving software systems that people can use. There have been a number of educational interventions developed to teach people how to thwart various cyber attacks such as phishing [3]. Surprisingly, no previous study has given sufficient consideration to develop training or educational interventions to enhance software developers' secure coding behaviour. Therefore, this research focuses on designing a serious game design framework as an educational tool for software developers' to improve their (secure) coding behaviour, so they can develop privacy-preserving software systems that people can use (Figure1).

## 3 A Serious Game Design Framework for Enhancing Secure Coding Behaviour

The aim of the proposed serious game design framework is to enhance software developers' (secure) coding behaviour through their motivation deriving elements from a previously invented game design framework for threat avoidance behaviour [2], Bloom's taxonomy [12] and data minimisation model [21].

Previous research has developed a systematic approach (i.e. methodology) that enabled software developers to make their decisions to minimise user data in software systems they develop through understanding data [21]. Authors claimed that their methodology encourages software developers to think of the ways they would use data in a system design, focusing on the collection, storage and sharing of data. Therefore, the proposed game design framework in this research teaches how software developers make their decisions to minimise user data when developing privacy-preserving software systems through the invented data minimisation model [21]. This improves software developers' (secure) coding behaviour through their motivation.

Bloom's taxonomy [12], which explains process of learning, is used to incorporate the data minimisation model into the game framework that teaches software developers to embed privacy into software systems they develop. Furthermore, it demonstrates a classification of levels of intellectual behaviour important in learning [1].

The proposed game design framework explains how software developers can improve their (secure) coding behaviour through motivation, so they can develop privacy-preserving software systems that people can use (Figure1) [2,3,12].

Consistent with the serious game design framework (Figure1), software developers' (secure) coding behaviour is determined by motivation, which, in turn, is affected by their threat perception (i.e. Perceived Threat). Perceived threat is influenced by perceived severity and susceptibility. Threat perception of software developers (i.e. perceived threat) is also influenced by the interaction effect of perceived severity and susceptibility. Software developers' motivation of writing a secure code snippet is also determined by three constructs such as safeguard effectiveness, safeguard cost and self-efficacy.

Whilst the game design framework identifies the issues that the game design needs to address, it should also indicate how to structure this information and present it in a gaming context. To this end, we aimed to develop threat perceptions, making software developers more motivated to write a secure code snippet through the proposed serious game design framework.

Therefore, the hypotheses (H1...H7) developed in the proposed game design framework are described in the context of enhancing software developers' secure coding behaviour as follows:

**H1.** Motivation positively affects software developers' secure coding behaviour to develop privacy-preserving software systems.

**H2.** Perceived threat positively affects software developers' motivation to develop privacy-preserving software systems.

**H3a.** Perceived severity positively affects software developers' threat perception when developing privacy-preserving software systems.

**H3b.** Perceived susceptibility positively affects software developers' threat perception when developing privacy-preserving software systems.

**H3c.** The combination of perceived severity and perceived susceptibility positively affects software developers' threat perception when developing privacy-preserving software systems.

**H4.** Safeguard effectiveness positively affects software developers' motivation to develop privacy-preserving software systems.

**H5.** Safeguard cost negatively affects software developers' motivation to develop privacy-preserving software systems.

**H6.** Self-efficacy positively affects software developers' motivation to develop privacy-preserving software systems.

**H7.** Process of learning (i.e. data minimisation technique [21]) through the Bloom's taxonomy [12] positively affects software developers' self-efficacy to develop privacy-preserving software systems.

## 4 Application Scenario

The aim of the proposed game framework is to educate software developers to write a better code snippet enhancing their secure coding behaviour, so they can develop privacy-preserving software systems that people can use. To achieve this, the proposed serious game design framework derived the elements from a previously invented game design framework [2] for threat avoidance behaviour, Bloom's taxonomy [12] and data minimisation model [21]. Therefore, this section describes how one can incorporate the elements of the proposed game framework into a possible application scenario (i.e. a game design scenario) enabling developers to enhance their (secure) coding behaviour through motivation.

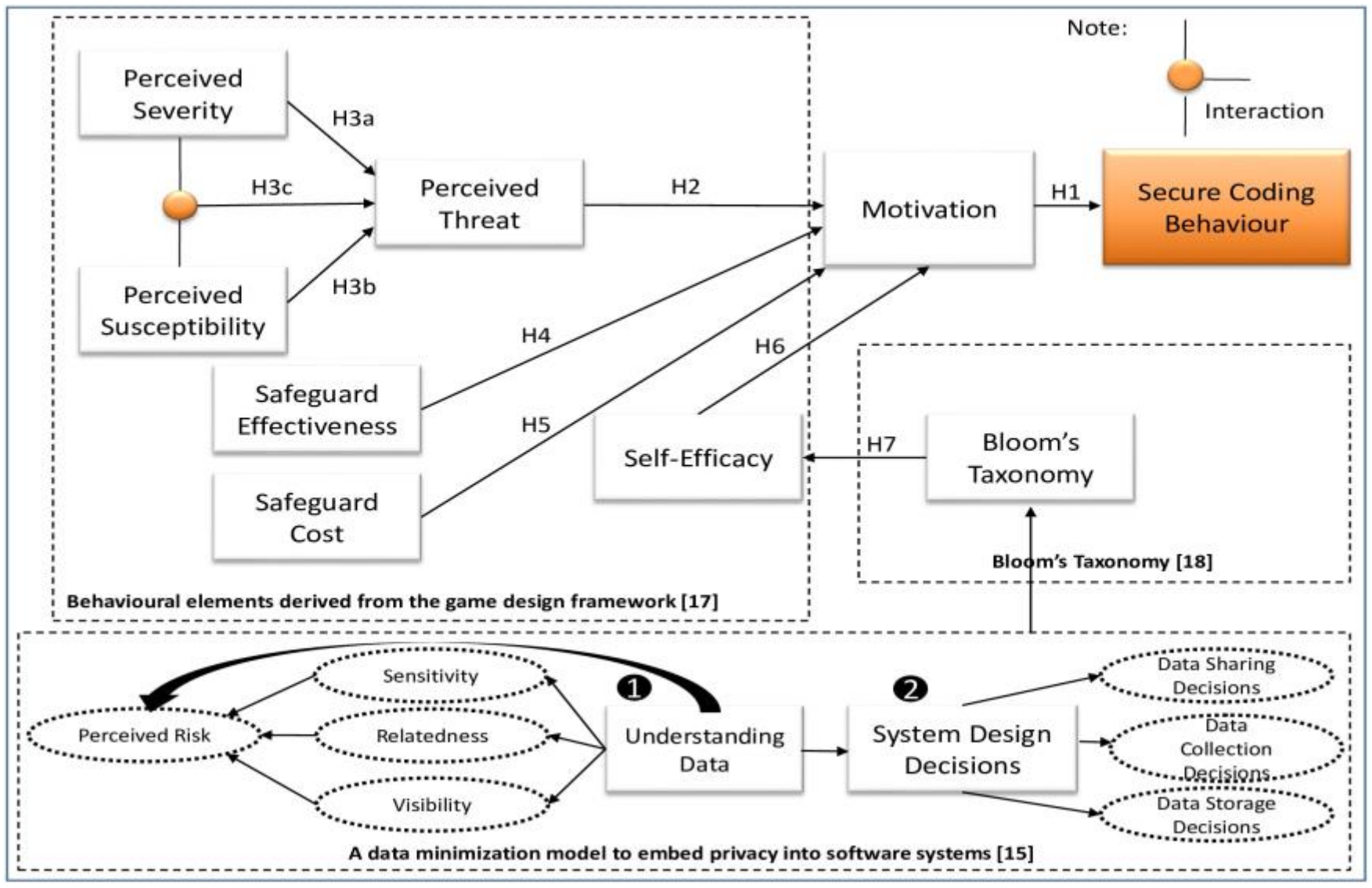

**Figure 1.** A serious game design framework for enhancing software developers' secure coding behaviour (elements from the game design framework [2], Bloom's taxonomy [12] and the data minimisation model [21])

The proposed game design scenario is developed based on a story that simplifies and exaggerates real life. Software developer role-plays as the game player in the game design scenario. In the beginning, a gaming environment can be presented with a game player depicting some scenarios related to data breaches, perhaps due to poorly written software systems through demo video clips. Therefore, the game player's target is to fix the issue(s) in the software system that caused data breaches. This will not only present the privacy threat perception (*Perceived Threat*) but also it shows the importance of using Data Minimisation (DM) technique (i.e. minimising user data in systems) when developing software applications that preserve user privacy.

Data breach in a software application can occur in several ways [27]. For example, weak and stolen credentials (a.k.a. passwords), application vulnerabilities, malware, social engineering, too many permissions, insider threats, physical attacks, improper configuration or user error. The gaming story and environment should be designed as a means, where the player *perceives a privacy threat* that the user data of ap- plications they develop can be susceptible (*Perceived Suscep- tibility*) in a data breach by any means (e.g. weak and stolen credentials, social engineering, etc.). The game player should also perceive through the designed game if the particular data breach occurs through the application he develops, what would be the severity (i.e. negative consequences) looks like, for example, financial loss or reputation damage (*Perceived Severity*). Therefore, developing a privacy *threat perception* in the game player's mind demonstrating perceived severity and susceptibility, will motivate himself to enhance coding behaviour, contributing to developing privacy-preserving applications.

The game player (i.e. software developer) should also be able to develop his confidence (i.e. *Self-Efficacy*) as they learn how software developers make their decisions to minimise user data when developing software systems that preserve user privacy. Self-Efficacy has a correlation with the one's knowledge [1, 13]. For example, when software developers are knowledgeable of how to minimise user data, they are more motivated and confident in taking relevant actions to develop privacy-preserving software systems. Therefore, the game player is taught how to "understand data" before "making systems design decisions" (on minimising the data use in the system) through the developed data minimisation model [21]. A calculation model developed through an empirical investigation to measure the perceived "data" privacy risk is used to understand data [20] (i.e. *Understanding Data* shown in Figure1). "Perceived privacy risks" of data is measured from the user perspective (Figure1): through the "sensitivity" of data to the user, "visibility" of data in the system, and the "relatedness" of the data to the system as parameters [20]. Privacy risk of data, the user perceives $P_{i,j}$ when disclosing a data element $D_i$ in an application context $C_j$ as,

Perceived Privacy Risk

$$P_{(i,j)} = \alpha * \frac{S^a_{(i,j)} \times V^b_{(i,j)}}{R^c_{(i,j)}}$$

where a, b, c and $\alpha$ values could take any real number. However, to avoid any confusion, this research uses an approximation for a, b, c limited to whole numbers.

According to this calculation, the Privacy Risk $P_{(D_i)}$ of a data element $D_i$ in an application context can be defined by $C_j \in \{x \mid x \in \mathbb{R}$ where, $0 < x\}$.

As the model calculated [20], sensitivity implies the impact of loss of a particular data item to the user [14]. For example, for a user, losing the credit card number may have a higher impact than losing his date of birth. It captures this variation of data by sensitivity. Visibility is described as a measure of how widely available (i.e. visible) the data item becomes in the system [14]. For example, one's "privateness" on Facebook is affected by the visibility of some data elements [15], such as hiding someone's relationship status from the general public. The visibility of data hence is determined by the software developer and captures this variation. Relatedness measures how relevant the data item is being collected for the purpose of the software application. It captures how relevant the data item is for the software application by the relatedness parameter.

Therefore, to understand the data through the developed model [20], the game player (i.e. software developer) is required to determine the following items in a given scenario presented in the game design, a. How sensitive data would be for a user (highly sensitive, moderately sensitive, low sensitive)? b. How visible the data would be in the system (highly visible, moderately visible, low visible)? and c. How relevant the data is to the system (highly relevant, moderately relevant, low relevant)?

In the game design, the player (i.e. software developer) needs to scale data against each parameter such as sensitivity, severity and relatedness [20]. This encourages software developers to think of each data item they use in the system, its purpose and whether or not that particular data item should be in the system when designing. On the other hand, this developer thinking ensures if they must collect unnecessary data in the system or what countermeasures should be in place to protect data that is more sensitive. Therefore, understanding these parameters (i.e. sensitivity, visibility and relatedness shown in Figure 1) of data and measuring the "data" privacy risk using them, enable the developers' decision making capability on how they should protect these data items (i.e. when collecting, storing and sharing) when developing software systems that preserve user privacy.

Moreover, previous research emphasises the importance of knowledge management to support creation, transfer and application of knowledge in a particular context [1]. Therefore, Bloom's taxonomy [12] is used to incorporate the aforementioned teaching content developed through data minimisation model [21] into the game design framework enabling the game player to have more confidence (*Self-Efficacy*) of devel- oping privacy-preserving software applications. Following the steps of Bloom's taxonomy [12] embedded in the game design scenario, a game player is able to enhance (secure) coding behaviour. Therefore, the game application scenario is designed incorporating the teaching content from the data minimisation model [20] using Bloom's taxonomy [12] as follows:

**a. Remembering:** The game player (i.e. software developer) is able to recall or remember information (e.g. understanding data and making system design decisions) s/he learnt through the game.

**b. Understanding:** The game player is able to explain ideas or concepts (e.g. understanding data, making system design decisions, measuring perceived privacy risk of data) through the game.

**c. Applying:** The game player is able to apply what s/he learnt from the game (e.g. understanding data, making system design decisions, measuring perceived privacy risk of data) when developing various software applications that preserve user privacy.

**d. Analysing:** The game player is able to distinguish between the different parts of what s/he has learnt through the game. For example, initially understanding data (i.e. perceived privacy risk) and then making system decisions

**e. Evaluating:** The game player is able to justify a stand or decision on the developed privacy-preserving software systems.

**f. Creating:** The game player is able to develop new processes, models or point of view that can be used to develop new software systems that preserve user privacy.

In summary, the elements of the game design framework (Figure 1) were incorporated into a serious game application as an educational tool for software developers to improve (secure) coding behaviour, so they can develop privacy-preserving software applications.

## 5 Conclusion and Future work

This research proposed a serious game design framework as an educational tool for software developers to improve (secure) coding behaviour, so they can develop privacy-preserving software applications that people can use. To achieve this, we derived the elements from a previously invented game design framework [2], Bloom's taxonomy [12] and data minimisation model [21]. The elements derived from the previous invented game design framework focused on developing a "privacy" threat perception. In addition, it also nudges the player to perceive the importance of developing privacy-preserving software systems that can protect user privacy. The teaching content (i.e. how to minimise data when designing systems) derived from the data minimisation model [20, 21] incorporated into a game application scenario in a way that the player can learn and develop his confidence. To present the teaching content to the game player as an interactive means, for example, to support creation, transfer and application of knowledge in a privacy-preserving application development context, we used the elements/steps provided in Bloom's taxonomy [12]. Therefore, the player is more confident (i.e.

*Self-Efficacy*) and even willing take relevant actions to de- velop their software systems that preserve user privacy.

Further research should be undertaken to empirically investigate the proposed game design framework through software developers to examine their (secure) coding behaviour. Gaming prototypes can be designed (i.e. through story-boarding) and developed (i.e. using both low and high fidelity application prototypes) through various design techniques such as participatory or a scenario based game design. An experimental protocol such as a think-aloud study, can be employed to examine the participants' impact on the developed game design framework (i.e. secure coding behaviour) after their engagement with the developed game prototype.

## Acknowledgments

We would like to thank Adam Shostack (President of Shostack & Associates, a consultancy in Seattle, WA, USA) for his feedback on our initial manuscript draft.

## References

[1] Aisa Amagir, Wim Groot, Henriëtte Maassen van den Brink, and Arie Wilschut. 2020. Financial literacy of high school students in the Netherlands: knowledge, attitudes, self-efficacy, and behavior. *International Review of Economics Education* (2020), 100185.

[2] Nalin Asanka Gamagedara Arachchilage and Steve Love. 2013. A game design framework for avoiding phishing attacks. *Computers in Human Behavior* 29, 3 (2013), 706–714.

[3] Nalin Asanka Gamagedara Arachchilage, Steve Love, and Konstantin Beznosov. 2016. Phishing threat avoidance behaviour: An empirical investigation. *Computers in Human Behavior* 60 (2016), 185–197.

[4] Oshrat Ayalon, Eran Toch, Irit Hadar, and Michael Birnhack. 2017. How developers make design decisions about users' privacy: the place of professional communities and organizational climate. In *Companion of the 2017 ACM Conference on Computer Supported Cooperative Work and Social Computing*. ACM, 135–138.

[5] Fei Bu, Nengmin Wang, Bin Jiang, and Huigang Liang. 2020. "Privacy by Design" implementation: Information system engineers' perspective. *International Journal of Information Management* 53 (2020), 102124.

[6] Carole Cadwalladr and Emma Graham-Harrison. 2020. Revealed: 50 million Facebook profiles harvested for Cambridge Analytica in major data breach. https://www.theguardian.com/news/2018/mar/17/cambridge-analytica-facebook-influence-us-election. [Online: Accessed 29-May-2020].

[7] Ann Cavoukian. 2009. Privacy by Design, The answer to overcoming negative externalities arising from poor management of personal data. In *Trust Economics Workshop London, England, June*, Vol. 23. 2009.

[8] Gabriel Alberto García-Mireles, Miguel Ehécatl Morales-Trujillo, Mario Piattini, and Erick Orlando Matla-Cruz. 2019. A Systematic Mapping Study on Privacy by Design in Software Engineering. (2019).

[9] Irit Hadar, Tomer Hasson, Oshrat Ayalon, Eran Toch, Michael Birnhack, Sofia Sherman, and Arod Balissa. 2017. Privacy by designers: software developers' privacy mindset. *Empirical Software Engineering* (2017), 1–31.

[10] Shubham Jain and Janne Lindqvist. 2014. Should I protect you? Understanding developers' behavior to privacy-preserving APIs. In *Proceed- ings of the Workshop on Usable Security (USEC 2014). Internet Society*.

[11] Pawel Kamocki and Andreas Witt. 2020. Privacy by Design and Language Resources. In *Proceedings of The 12th Language Resources and Evaluation Conference*. 3423–3427.

[12] David R Krathwohl and Lorin W Anderson. 2009. *A taxonomy for learning, teaching, and assessing: A revision of Bloom's taxonomy of educational objectives*. Longman.

[13] Youngsun Kwak, Seyoung Lee, Amanda Damiano, and Arun Vishwanath. 2020. Why Do Users Not Report Spear Phishing Emails? *Telematics and Informatics* (2020), 101343.

[14] E Michael Maximilien, Tyrone Grandison, Tony Sun, Dwayne Richardson, Sherry Guo, and Kun Liu. 2009. Privacy-as-a-service: models, algorithms, and results on the Facebook platform. In *Proceedings of Web*, Vol. 2.

[15] Tehila Minkus and Nasir Memon. 2014. On a scale from 1 to 10, how private are you? Scoring Facebook privacy settings. In *Proceedings of the Workshop on Usable Security (USEC 2014). Internet Society*.

[16] Marie Caroline Oetzel and Sarah Spiekermann. 2014. A systematic methodology for privacy impact assessments: a design science approach. *European Journal of Information Systems* 23, 2 (2014), 126–150.

[17] Ashwini Rao, Florian Schaub, Norman Sadeh, Alessandro Acquisti, and Ruogu Kang. 2016. Expecting the unexpected: understanding mismatched privacy expectations online. In *Symposium on Usable Privacy and Security (SOUPS)*.

[18] Jeff Sedayao, Rahul Bhardwaj, and Nakul Gorade. 2014. Making big data, privacy, and anonymization work together in the enterprise: experiences and issues. In *Big Data (BigData Congress), 2014 IEEE International Congress on*. IEEE, 601–607.

[19] Awanthika Senarath and Nalin AG Arachchilage. 2018. Why developers cannot embed privacy into software systems? An empirical investigation. In *Proceedings of the 22nd International Conference on Evaluation and Assessment in Software Engineering 2018*. 211–216.

[20] Awanthika Senarath and Nalin AG Arachchilage. 2019. A model for system developers to measure the privacy risk of data. In *Proceedings of the 52nd Hawaii International Conference on System Sciences*.

[21] Awanthika Senarath and Nalin Asanka Gamagedara Arachchilage. 2019. A data minimization model for embedding privacy into software systems. *Computers & Security* 87 (2019), 101605.

[22] Awanthika Senarath, Marthie Grobler, and Nalin Asanka Gamagedara Arachchilage. 2019. Will they use it or not? Investigating software developers' intention to follow Privacy Engineering Methodologies (PEMs). *ACM Transactions on Privacy and Security (TOPS)* 22, 4 (2019), 23.

[23] Swapneel Sheth, Gail Kaiser, and Walid Maalej. 2014. Us and them: a study of privacy requirements across North America, Asia, and Europe. In *Proceedings of the 36th International Conference on Software Engineering*. ACM, 859–870.

[24] Daniel J Solove and Paul Schwartz. 2014. *Information privacy law*. Wolters Kluwer Law & Business.

[25] Christine Utz, Martin Degeling, Sascha Fahl, Florian Schaub, and Thorsten Holz. 2019. (Un) informed Consent: Studying GDPR Consent Notices in the Field. In *Proceedings of the 2019 ACM SIGSAC Conference on Computer and Communications Security*. 973–990.

[26] Charles Weir, Ben Hermann, and Sascha Fahl. 2020. From Needs to Actions to Secure Apps? The Effect of Requirements and Developer Practices on App Security. In *29th {USENIX} Security Symposium ({USENIX} Security 20)*.

[27] Chamila Wijayarathna, Marthie Grobler, and Nalin AG Arachchilage. 2019. Software developers need help too! Developing a methodology to analyse cognitive dimension-based feedback on usability. *Behaviour & Information Technology* (2019), 1–22.

[28] Glenn Wurster and Paul C van Oorschot. 2009. The developer is the enemy. In *Proceedings of the 2008 workshop on New security paradigms*. ACM, 89–97.